\documentclass[conference]{IEEEtran}
\IEEEoverridecommandlockouts
\usepackage[utf8]{inputenc}
\usepackage{balance} %
\usepackage{amsmath,amsfonts}
\usepackage{algorithmic}
\usepackage{graphicx}
\usepackage{caption}
\usepackage{textcomp}
\usepackage{xcolor}
\usepackage{filecontents} 
\def\BibTeX{{\rm B\kern-.05em{\sc i\kern-.025em b}\kern-.08em
    T\kern-.1667em\lower.7ex\hbox{E}\kern-.125emX}}
\usepackage[textsize=tiny, color=green!50]{todonotes}
\usepackage[nolist]{acronym} %
\usepackage{tikz} %
\usetikzlibrary{positioning} %
\usepackage{booktabs}
\usepackage{multirow}
\usepackage{colortbl}
\usepackage{pgfkeys}
\usepackage[cachedir=./mintedcache, frozencache=true]{minted}
\usepackage{minted}
\usepackage{verbatim}
\usepackage{url}
\usepackage{meta/lstcustom}
\usepackage{caption}
\usepackage{subcaption}
\usepackage{csvsimple} %
\usepackage{mdframed}
\usepackage{tcolorbox}
\tcbuselibrary{theorems}
\usepackage{wasysym} %
\usepackage{tcolorbox}
\AtBeginEnvironment{tcolorbox}{\small}
\usepackage{tablefootnote}
\usepackage{cleveref}
\usepackage{enumitem}
\usepackage{xurl} %

\newcommand{\ccsast}{\textit{CogniCrypt\textsubscript{SAST}}}

\newcommand{\code}[1]{\mintinline{java}{#1}}
\newcommand{\class}[1]{\mintinline{java}{#1}}
\newcommand{\errtype}[1]{\emph{#1}}
\newcommand{\error}[1]{\emph{#1}}

\definecolor{gray}{gray}{0.85}
\definecolor{light-gray}{gray}{0.9}
\definecolor{lgray}{gray}{0.9}
\definecolor{firebrick}{rgb}{0.7, 0.13, 0.13}
\definecolor{mountainmeadow}{rgb}{0.19, 0.73, 0.56}
\definecolor{viridisblue}{RGB}{59, 82, 139}
\definecolor{viridisgreen}{RGB}{37, 172, 130}

\newcommand{\cover}{{\color{black!70!black}{\Circle}}}
\newcommand{\nocover}{{\color{viridisgreen}{\CIRCLE}}}
\newcommand{\err}{{\color{viridisblue}{\CIRCLE}}}
\newcommand{\noerr}{\cover}

\lstdefinelanguage{CrySL}[]{Java}{
  morekeywords={ABSTRACT, SPEC, OBJECTS, EVENTS, ORDER, CONSTRAINTS, REQUIRES, ENSURES, REFINES, define, add, constraint, SEVERITY, WHEN, from, to},
  moredelim=[is][\textcolor{darkgray}]{\%\%}{\%\%},
  moredelim=[il][\textcolor{darkgray}]{§§}
}

\newtcbtheorem{obs}{Research Question}%
{colback=viridisblue!5,colframe=viridisblue,fonttitle=\bfseries}{th}

\makeatletter %
\newcommand{\linebreakand}{%
  \end{@IEEEauthorhalign}
  \hfill\mbox{}\par
  \mbox{}\hfill\begin{@IEEEauthorhalign}
}
\makeatother %

\def\BibTeX{{\rm B\kern-.05em{\sc i\kern-.025em b}\kern-.08em
    T\kern-.1667em\lower.7ex\hbox{E}\kern-.125emX}}
\begin{document}

\title{To Fix or Not to Fix: A Critical Study of Crypto-misuses in the Wild}
\author{\IEEEauthorblockN{Anna-Katharina Wickert\IEEEauthorrefmark{1}, Lars Baumgärtner\IEEEauthorrefmark{1}, Michael Schlichtig\IEEEauthorrefmark{2}, Krishna Narasimhan\IEEEauthorrefmark{1}, Mira Mezini\IEEEauthorrefmark{1}} 
\IEEEauthorblockA{\IEEEauthorrefmark{1}Technische Universität Darmstadt\\
Darmstadt, Germany\\
Email: <wickert,baumgaertner,kri.nara,mezini>@cs.tu-darmstadt.de, } \IEEEauthorblockA{\IEEEauthorrefmark{2}Heinz Nixdorf Institute, Paderborn University \\
Paderborn, Germany \\
Email: michael.schlichtig@uni-paderborn.de}}

\IEEEoverridecommandlockouts
\IEEEpubid{\makebox[\columnwidth]{\copyright2022 IEEE \hfill} \hspace{\columnsep}\makebox[\columnwidth]{ }}
\maketitle

\IEEEpubidadjcol
\pgfkeyssetvalue{projects_all}{210}
\pgfkeyssetvalue{projects_all}{210}
\pgfkeyssetvalue{projects_jca}{210}
\pgfkeyssetvalue{projects_bc}{7}
\pgfkeyssetvalue{objects_all}{3,294}
\pgfkeyssetvalue{objects_jca}{3,223}
\pgfkeyssetvalue{objects_bc}{71}
\pgfkeyssetvalue{misuses_all}{2,695}
\pgfkeyssetvalue{misuses_jca}{2,634}
\pgfkeyssetvalue{misuses_bc}{61}
\pgfkeyssetvalue{misuses_all_project}{185}
\pgfkeyssetvalue{misuses_jca_project}{185}
\pgfkeyssetvalue{misuses_bc_project}{5}
\pgfkeyssetvalue{misuses_jca_project_perc}{88.10}
\pgfkeyssetvalue{misuses_bc_project_perc}{71.43}
\pgfkeyssetvalue{projects_only_insec}{82}
\pgfkeyssetvalue{projects_only_insec_jca}{80}
\pgfkeyssetvalue{projects_only_insec_bc}{2}
\pgfkeyssetvalue{required_total}{960}
\pgfkeyssetvalue{incomplete_total}{815}
\pgfkeyssetvalue{constraint_total}{566}
\pgfkeyssetvalue{topfourclasses_perc}{61.86}

\pgfkeyssetvalue{tp}{93}
\pgfkeyssetvalue{fp}{33}
\pgfkeyssetvalue{idk}{31}
\pgfkeyssetvalue{non-sec-misuses}{9}
\pgfkeyssetvalue{non-sec-misuses-md}{8}
\pgfkeyssetvalue{non-sec-misuses-md_perc}{22.86}
\pgfkeyssetvalue{tp-grep}{25}
\pgfkeyssetvalue{tp-order}{21}
\pgfkeyssetvalue{tp-string}{22}
\pgfkeyssetvalue{tp-default}{15}
\pgfkeyssetvalue{tp-hardcoded}{2}
\pgfkeyssetvalue{loop-fp}{2}
\pgfkeyssetvalue{loop}{8}
\pgfkeyssetvalue{fp-crysl}{17}
\pgfkeyssetvalue{fp-model}{11}
\pgfkeyssetvalue{fp-decrypt}{1}

\begin{abstract}
Recent studies 
have revealed that 87~\% to 96~\% of the Android apps using cryptographic APIs have a misuse
which may cause security vulnerabilities. 
As previous studies did not conduct a qualitative examination of the validity and severity of the findings, our objective was to understand the findings in more depth. 
We analyzed a set of 936 open-source Java applications for cryptographic misuses. %
Our study reveals that \pgfkeysvalueof{misuses_jca_project_perc}~\% of the analyzed applications fail to use cryptographic APIs securely. 
Through our manual analysis of a random sample, we
gained new insights into \textit{effective false positives}. 
For example, every fourth misuse of the frequently misused JCA class \class{MessageDigest} is an \textit{effective false positive} due to its occurrence in a non-security context. 
As we wanted to gain deeper insights into the security implications of these misuses, 
we created an extensive vulnerability model for cryptographic API misuses.
Our model includes previously undiscussed attacks in the context of cryptographic APIs such as DoS attacks.
This model 
reveals
that nearly half of the misuses are of high severity, e.g., hard-coded credentials and potential Man-in-the-Middle attacks. %

\end{abstract}


\setminted{
    numbersep=3pt,
    xleftmargin=12pt,
}

\section{Introduction}
\label{sec:intro}

Many applications need to protect sensitive and high-value data, such as passwords or financial transactions, 
using cryptography (hereafter referred to as \textit{crypto}). 
For this purpose, crypto APIs provide access to crypto tasks, protocols, and primitives in all major programming languages. 
However, several studies have revealed that developers struggle to use crypto APIs~\cite{lazar_why_2014,nadi_jumping_2016} correctly and introduce misuses, meaning that an API usage may be 
used syntactically correct but problematic from a security perspective.
A common crypto misuse is the use of the \textit{Electronic Code Book (ECB)} mode for encryption.
Although it has been known for a long time that 
\textit{ECB} is insecure~\cite{egele_empirical_2013}, it was, e.g., 
found in 
the widely used Zoom video conferencing system
until May 2020\footnote{\url{https://support.zoom.us/hc/en-us/articles/360043770412-Updating-your-Zoom-Rooms-to-version-5-0-5}}.

To address this problem, static analyzers such as \textit{SpotBugs}\footnote{https://spotbugs.github.io/}, 
\textit{CryptoGuard}~\cite{rahaman_cryptoguard:_2019}, and \ccsast{}~\cite{kruger2019crysl}, have been proposed to support developers and security researchers in checking for such misuses.
These tools have been used in various empirical studies that have produced worrying insights into the state of crypto usages in the wild \cite{egele_empirical_2013, kruger2019crysl, rahaman_cryptoguard:_2019, gao2019negative}.
However, none of the studies performed a qualitative examination of the validity and severity of the findings. 
Such an examination is, however, essential for both getting a more realistic picture of the state of crypto usages in the wild and %
to improve future studies and analyzes.
It is well-known that static analyzes may produce false positives, which are considered their \emph{Achilles heel}~\cite{johnson_dev, wagner2005comparing}. 
Moreover, in a security context, the validity of any finding should be judged from the perspective of a threat model to get a feeling for its severity. 

This paper contributes to closing this gap by designing and conducting a study focusing on qualitative examination of reported crypto misuses. 
With regard to false positives, we are particularly interested in reports that are specific to their concrete usage. 
An API may be used in different – non-crypto - contexts and not the same constraints apply in all contexts. 
For example, the most misused JCA class \code{MessageDigest} from previous studies~\cite{gao2019negative,kruger2019crysl,rahaman_cryptoguard:_2019} may be used to compute hashes independent of a security context, e.g., to compute the hash of a file. 
However, a static analysis cannot know this and has to be conservative. 
Thus, it may produce false positives and draw an inaccurate picture of the security of our software. 
With regard to qualitatively judging the severity of findings, we formulate and use a novel, comprehensive threat model.
Specifically, we designed and conducted a study to answer the following research questions:
\begin{enumerate}
    \item[RQ1:] What are common (effective) false positives arising from misuses, e.g., due to a non-security context?
    \item[RQ2:] How severe are the vulnerabilities introduced by crypto API misuses in applications?
\end{enumerate}

We studied crypto misuses
of two Java crypto providers -- the Java Cryptography Architecture (JCA) and Bouncy Castle (BC) -- in open-source Java projects from GitHub
using \ccsast{}~\cite{kruger2019crysl}. 
We only included projects that are mainly developed and maintained by professionals to address the generalizability problem of open-source studies~\cite{spinellis_dataset_2020}.
Altogether, we collected
936 Java projects, of which \pgfkeysvalueof{projects_all} use a crypto API and \pgfkeysvalueof{misuses_jca_project_perc}~\% have at least one misuse.

To qualitatively analyze the misuses,
we randomly picked 157 misuses
for review.
We observed that one fourth of the misuses of the class \class{MessageDigest} -- the most misused class according to previous studies~\cite{kruger2019crysl, rahaman_cryptoguard:_2019, gao2019negative} and the second most in our study -- occur in a non-security context.
Thus, we can consider these misuses as \textit{effective false positives}~\cite{sadowski2015Tricorder}
that may not or cannot be fixed by developers.

To judge the severity of all findings, we defined a threat model that connects API misuses reported by \ccsast{} to security vulnerabilities. 
This model subsumes the existing vulnerability model for crypto API misuses by Rahaman et al.~\cite{rahaman_cryptoguard:_2019} and includes new threats, e.g., \textit{Denial of Service} (DoS) attack and \textit{Chosen-Ciphertext Attacks} (\textit{CCA}). %
Overall, our model marks 42.78~\% of the 
misuses as high severity.

In summary, this paper makes the following contributions:
\begin{itemize}
    \item We studied crypto misuses found in a large, representative data set of applications~\cite{spinellis_dataset_2020}. 
    \item We provide empirical evidence that the reported misuses contain a significant amount of \textit{effective false positives}. 
    Our disclosure confirms this and reveals that many may not be fixed even though some are ranked as high severity. %
    \item We created a novel, comprehensive threat model mapping crypto API misuses to vulnerabilities, introducing previously undiscussed threats like DoS attacks and CCAs.%
\end{itemize}

\section{Background}
\label{sec:backgroundAndRelatedWork}
\label{sec:background}
\begin{listing}
\caption{Code snippet that signs a byte array using a key.}

\label{lst:pdkdf2Encrypt}
\label{lst:background}
\begin{minted}[breaklines,numbers=left,firstnumber=last,fontsize=\small,escapeinside=!!]{java}
private byte[] signByte(byte[] dataToSign){
   byte[] signedBytes;
   Signature s = Signature.getInstance("SHA1WithRSA");!\label{lst:constraint}!
   s.initSign(getPrivateKey()); !\label{lst:required}!
   s.update(dataToSign); !\label{lst:l:update}!
   // Call to signedBytes = s.sign() missing.  !\label{lst:l:incompleteoperation}!
   return signedBytes;
}

// Get a PrivateKey object with an insecure key length.
private PrivateKey getPrivateKey() {
    return shortKey();
}
\end{minted}
\end{listing}

In this section, we provide background information regarding crypto API misuses in Java and introduce \ccsast{} \cite{kruger2019crysl}, the crypto API misuse analyzer we used for our study.
In addition, we introduce the term \textit{effective false positives}. 

\subsection{Misuses of Java Crypto APIs}
\label{sec:cryptomisuses}
\label{sec:background:misuses}

The JCA provides a set of extensible cryptographic components ranging from encryption over authentication to access control, enabling developers to secure their applications. 
It is implementation-independent by using a "provider" architecture; developers can plug and play their implementation of crypto primitives for use with this architecture. 
A commonly used provider besides the default that is shipped with the Java Development Kit (JDK) is the Bouncy Castle library. 

Listing~\ref{lst:background} illustrates a usage of the JCA to sign a byte-array \mintinline{Java}{dataToSign}. 
For this, the \code{Signature} object is initialized with a signature algorithm (Line~\ref{lst:constraint}) and a private key, passed via the function \code{getPrivateKey}, is added to the \code{Signature} object \code{s} (Line~\ref{lst:required}).
Next, the byte-array \code{dataToSign} is passed to \code{s} to actually compute the signature of the data (Line~\ref{lst:l:update}) before the function returns the signed bytes. 
Unfortunately, the call to \code{sign} that would return the signature is missing (Line~\ref{lst:l:incompleteoperation}).

A \emph{crypto misuse}, hereafter just misuse, is a usage of a crypto API that is considered insecure by experts. 
A misuse may be syntactically correct, a working API usage, and may not even raise an exception. 
We will briefly discuss the error types defined by Krüger et al.~\cite{kruger2019crysl} to illustrate some misuses. 

\begin{enumerate}
    \item \textbf{Constraint Errors} (Listing~\ref{lst:background}, Line~\ref{lst:constraint}): Crypto APIs use parameters to let developers select crypto algorithms when initializing crypto objects.
    These parameters are often passed as strings 
    in a specific format. 
    
    \item \textbf{Incomplete Operation Errors} (Listing~\ref{lst:background}, Line~\ref{lst:l:incompleteoperation}):
    The security of crypto objects may rely on a specific protocol. %
    For instance, \code{Signature} crypto objects, once initialized and filled with data via a call to \code{update(byte[])}, require a call to the \code{sign} method to complete the signing operation. 
    Thus, the required calls to complete the use of an initialized crypto object are missing. 
    
    \item \textbf{Required Predicate Errors} (Listing \ref{lst:background}, Line~\ref{lst:required}): Crypto objects often depend on each other. 
    For example, \texttt{Signature} objects require a correctly generated \texttt{Key} object. %
    In order for a composed crypto solution to be secure, it is required that its components 
    on which it depends are secure.
    Thus, composing a crypto object with required but insecure objects results in a misuse. 
    
    \item \textbf{Never Type of Error:} 
    Sensitive information, e.g., a secret key, should never be of type
    \texttt{java.lang.String}, as strings are considered insecure compared to mutable byte arrays.
    Strings are immutable and stay in memory until collected by Java's garbage collector. 
    Thus, they are longer visible in memory for attackers than necessary and outside of the direct control of the developer~\footnote{\label{fn:jcastring}\url{https://docs.oracle.com/en/java/javase/17/docs/api/java.base/javax/crypto/spec/PBEKeySpec.html}, accessed 19.09.2022}.
    
    \item \textbf{Forbidden Method Errors:} Certain methods of crypto objects should never be called for security reasons. 
    For example, the \code{PBEKeySpec} object generates keys from passwords and requires a crypto salt while initializing the object. 
    Therefore, constructors that are not parameterized with a salt 
    cause a misuse.
    
    \item \textbf{Type State Error:} Such an error occurs when an object moves into an insecure state as the result of an improper method call sequence. 
    For example, a \code{Signature} object requires a call to the \code{initSign} method prior to any number of calls to the \code{update} method.  
    
    The key difference to an \error{Incomplete Operation Error} is that the missing and expected call is within the call sequence, while for an \error{Incomplete Operation Error} the expected call sequence is not finished. 
\end{enumerate}

\begin{table*}[t!]
    \footnotesize
    \centering
    \begin{tabular}{llllllllll}
        \toprule
        Vulnerabilities & Attack Type & Severity & Novel & C & IO & RP & NT & FM & TS \\ 
        \midrule
         Predictable/constant crypto keys &  \multirow{4}{*}{Predictability Through Initialization} & H & \cover  & \err & \noerr & \err & \noerr & \noerr & \err\\
         Predictable/constant passwords for PBE & & H & \cover & \noerr & \noerr & \noerr & \noerr &  \err & \noerr \\  
         Predictable/constant passwords for KeyStore & & H & \cover & \err & \noerr & \noerr  & \noerr & \noerr  &\noerr \\
         Cryptographically insecure PRNGs & & M & \cover & \noerr & \noerr & \err & \noerr & \noerr & \noerr \\
         \midrule
         Missed to finish crypto function &  \multirow{2}{*}{Predictability Through Usage} & H & \nocover & \noerr & \err & \noerr & \noerr & \noerr & \err \\
         Missed to pass data & & M & \nocover & \noerr & \err & \noerr & \noerr & \noerr & \err\\
         \midrule
         Insecure TrustManager & \multirow{2}{*}{MitM Attacks on SSL/TLS}  & H & \cover & \noerr & \noerr & \err & \noerr & \noerr & \noerr \\
         Insecure SSL/TLS standard & & H & \nocover & \err & \err & \noerr & \noerr & \err & \err \\
         \midrule 
         Static Salts in PBE & \multirow{3}{*}{Chosen-Plaintext Attack (CPA)} & M & \cover & \noerr & \noerr & \err & \noerr & \noerr & \noerr \\
         ECB mode in symmetric cipher & &M & \cover  & \err & \noerr & \noerr & \noerr & \noerr & \noerr \\
         Static IVs in CBC mode symmetric ciphers & & M & \cover & \noerr & \noerr & \err & \noerr & \noerr & \noerr \\
         \midrule
         Padding Oracle & Chosen-Aiphertext Attack (CCA) & M & \nocover & \err & \noerr & \noerr & \noerr & \noerr & \noerr \\
         \midrule
         Fewer than 10,000 iterations for PBE & \multirow{7}{*}{Bruteforce Attacks} & L & \cover & \err & \noerr & \noerr & \noerr & \noerr & \noerr \\
         64-bit block ciphers &  & L & \cover & \err & \noerr & \noerr & \noerr & \noerr & \noerr\\
         64-bit authentication tag GCM & & L & \nocover & \err & \noerr & \noerr & \noerr & \noerr & \noerr\\
         Insecure cryptographic ciphers & & L & \cover & \err & \noerr & \err & \noerr & \noerr & \noerr  \\
         Insecure cryptographic signature & & L & \nocover & \err & \noerr & \noerr & \noerr & \noerr & \noerr \\ 
         Insecure cryptographic MAC & & L & \nocover & \err & \noerr & \noerr & \noerr & \noerr & \noerr \\
         Insecure cryptographic hash & & H & \cover & \err & \noerr & \err & \noerr & \noerr & \noerr \\
         \midrule
         Usage of String & \multirow{2}{*}{Credential Dumping}  & L & \nocover & \noerr & \noerr & \noerr & \err & \noerr & \noerr \\
         Missed to clear password & & L & \nocover & \noerr & \err & \noerr & \noerr & \noerr & \noerr \\
         \midrule
         Trigger Exception & \multirow{1}{*}{DoS Attacks} & M & \nocover & \err & \noerr & \noerr & \noerr & \noerr & \err \\
         \bottomrule
    \end{tabular}
    \vspace{2mm}
    \captionsetup{justification=centering}
    \caption{A model of vulnerabilities which can be detected with \ccsast{}. 
    \footnotesize{For \underline{novel} \cover~marks if this vulnerability is discussed by Rahaman et al.~\cite{rahaman_cryptoguard:_2019} and \nocover~if not. 
    For \underline{C}: Constraint Error, \underline{IO}: Incomplete Operation Error, \underline{RP:} Required Predicate Error, \underline{NT:} Never Type of Error, \underline{FM:} Forbidden Method Error and \underline{TS:} Type State Error \err~marks if this vulnerability can be caused by the respective error type and \noerr~if not.}
    }
    \label{tab:vulnerabilities}
\end{table*}

\subsection{\ccsast{}}
\label{sec:app-ccsast}

For our study, we used the crypto misuse detector \ccsast{} which follows an allowlisting approach. 
In contrast to denylisting approaches such as \textit{CryptoGuard}~\cite{rahaman_cryptoguard:_2019} that describe vulnerabilities, allowlisting approaches describe all secure API usages. 
Violations of the 
defined rules are reported as misuses~\cite{kruger2019crysl}. 
Previous studies reported a precision of 85~\% to 94~\%~\cite{kruger2019crysl, hazhirpasand2020java}.
Further, \ccsast{} supports the BC library next to the JCA.

\subsection{Effective False Positives}
\label{def:efp}

Past research has shown that developers consider false positives as the "Achilles heel" of static analyzes~\cite{johnson_dev, ayewah2007evaluating}. 
However, in practice, the definition of false positives varies:
From a static analysis perspective, a false positive is a finding which is incorrectly identified by the analysis.
For developers on the other hand, some findings can not be fixed in the application, e.g., due to a broken standard. %
Sadowski et al.~\cite{sadowski2015Tricorder} introduce the term \textit{effective false positives} to cover reported misuses 
on which a user will not take further action. %

For an example of an \textit{effective false positive}, consider a usage of \textit{MD5} that is correctly flagged by the static analysis.
However, the usage of \textit{MD5} at hand happens in a non-security context as the concrete call cannot be influenced by external factors and is not essential for the security of the software.

\section{Threat-model of vulnerabilities introduced by API misuses}

\label{sec:vulnerabilititesmodel}
To reason about the potential impact of the reported misuses, we contribute a threat model extending existing models~\cite{rahaman_cryptoguard:_2019} with more vulnerabilities caused by API misuses. 
Specifically, we derive our model from a study of CrySL rules written by crypto experts, the respective APIs, and the misuses observed in our study. 
In combination with standard attacks and crypto misuses from previous work, our model covers a wide variety of crypto API misuses and their attack potential.
We list the vulnerabilities, the attack types, the respective severity, the novelty, and the affected error types in Table~\ref{tab:vulnerabilities}. %

\begin{enumerate}
    \item \textbf{Predictability Through Initialization:}
Applications can become insecure if sensitive information like a key is predictable~\cite{egele_empirical_2013, rahaman_cryptoguard:_2019, lazar_why_2014}. 
As an instance of this vulnerability, we consider predictable and constant crypto keys, e.g., a hard-coded key or predictable password used to derive a key from. %
Furthermore, a cryptographic secure random number in Java requires a non-predictable seed as well as the usage of a dedicated class, e.g., \texttt{java.security.SecureRandom}. %
If the application code misses to fullfill these requirements, the crypto operation becomes predictable. 
While Rahaman et al.~\cite{rahaman_cryptoguard:_2019} separated predictable keys and PRNGs, our model combines them, as both attack types are due to predictability. 

\item \textbf{Predictability Through Usage:}
In contrast to attacks of the type \textit{Predictability Through Initialization}, the improper usage of an API can render the respective computation predictable. 
An example of this issue is the usage of crypto APIs which rely on data to be processed, e.g., \class{MessageDigest}, and fail to process the required data. 
Thus, essentially resulting in a predictable computation. 

\item \textbf{MitM attacks on SSL/TLS:}
The improper usage of SSL/TLS can enable an attacker to launch a Man-in-the-Middle (MitM) attack to gain sensitive information~\cite{fahl_why_2012}. 
This includes improper configuration of connections, e.g., incorrect verification of protocols, as well as insecure cryptographic protocols, 
e.g., TLS 1.1,
which are vulnerable to attacks like the POODLE attack.%

\item \textbf{Chosen-Plaintext-Attack (CPA):} 
An encryption algorithm should be provably secure against chosen-plaintext-attacks (CPA)~\cite{egele_empirical_2013}. 
An example is a 
static \textit{Initialization Vector} (IV) for the \textit{Cipher Block Chaining} (CBC) mode. 

\item \textbf{Chosen-Ciphertext-Attack (CCA):}
While an encryption scheme should be provably secure against CPA, it should also be safe against chosen-ciphertext-attacks (CCA). 
Concretely, we cover improper padding schemes, like PKCS5 and PKCS7 in combination with the block cipher mode CBC~\cite{vaudenay2002security, klima2003side}, and the usage of plain RSA~\cite{bleichenbacher1998chosen} as these are all vulnerable to CCA. 
Note, that a padding attack requires an interaction between the attacker and the program that responds to messages from the attacker.
Thus, data at rest is not vulnerable. %

\item \textbf{Bruteforce Attacks:}
Some cryptographic algorithms are vulnerable to repeated, extensive computations, e.g., a feasible collision computation for MD5 and SHA-1~\cite{stevens2017first}.
Further, primitives like DES~\cite{rahaman_cryptoguard:_2019} with a block size of 64~bit or a 64-bit authentication tag for GCM~\cite{ferguson_2005} are vulnerable to brute-force attacks to break the encryption. 

\item \textbf{Credential dumping:}
In Java, string values are immutable and therefore cannot be cleared or overwritten from memory, except when the garbage collector runs, which is out of the developer's control. 
Thus, the JCA enforces the usage of byte arrays, e.g., \code{PBEKeySpec}%
, to handle sensitive information. 
However, developers may use strings to store this information and only convert it to byte arrays before calling the crypto API.
Furthermore, classes like \class{PBEKeySpec} provide the method \class{clearPassword} to clear the internal copy of the password. 
However, usages of this class without calling this method render an application vulnerable to credential dumping. 

\item \textbf{DoS Attacks:}
A Denial-of-Service (DoS) attack limits the availability of a service, e.g., by deliberately overloading the memory or consuming an extensive amount of CPU cycles.
One possibility to achieve this is by triggering an unintended behavior such as uncaught exceptions~\cite{wu2017uncaughtexceptions}.
Such exceptions can be raised when contradicting the API contract, e.g., by missing the initialization of a cipher object\footnote{\url{https://docs.oracle.com/en/java/javase/17/docs/api/java.base/javax/crypto/Cipher.html}, accessed 19.09.2022}. %
Depending on the program, such an exception can prevent further correct control flow,
can cause a program crash, or if triggered in rapid succession, it can spam the log file, increase CPU usage or cause IO congestion.
A recent example 
is CVE-2021-23372, where an exception causes the application to crash. %
\end{enumerate}

We categorize the severity into high, medium, and low.
Our prioritization is based on factors like whether the vulnerability can be exploited remotely, how difficult an attack is to perform, and if the attack leads to a direct gain or needs the combination with other flaws to be of use. 
Attacks such as MitM which can be triggered remotely and provide direct benefits for an attacker~\cite{rahaman_cryptoguard:_2019} are ranked as high severity. 
We mark vulnerabilities as medium severity, which lead to compromises of secrets for active, dedicated attackers.
These vulnerabilities are of great help in undermining the security of the systems~\cite{wijayarathna_using_2019}. 
Examples of this are CPA and CCA attacks.
An example of a vulnerability with low severity are credentials passed as a string, as these require prior access to the system or running process to be extracted. 
Thus, the attacker must have exploited other vulnerabilities before to gain access to the system. %

\section{Empirical Study}
\label{sec:study}

\label{sec:approach}
In this section, we describe the setup of our study and the insights gained from it. 
First, the
data set used for our analysis is described in Section~\ref{sec:setup}, followed by the 
misuses identified in Section~\ref{sec:eval-cc}.
Our research questions are answered in Section~\ref{sec:eval-manual} and \ref{sec:eval:vuln}, respectively. 
We published an artifact\footnote{\url{https://doi.org/10.6084/m9.figshare.21178243}} including our script for the threat model. %

\subsection{Dataset}
\label{sec:setup}
The dataset created by Spinellis et al.~\cite{spinellis_dataset_2020} addresses the generalizability problem of open-source studies and consists of 17,264 GitHub projects that are mainly developed or guided by enterprise employees and span multiple languages. 
\ccsast{} works on Java binaries, hence,
we filtered out the non-Java projects and attempted to compile the rest automatically for Maven/Gradle build instructions. 
This resulted in 936 Java enterprise-driven, open-source binaries, which serve as the basis of our study. 
We applied the \ccsast{} version 2.7.2\footnote{\url{https://github.com/CROSSINGTUD/CryptoAnalysis/releases/tag/2.7.2}} 
with the corresponding rule set for JCA and BC 
to them\footnote{Previous studies used older \ccsast{} versions \cite{kruger2019crysl,gao2019negative,rahaman_cryptoguard:_2019}.}. 
As the objective of our study was exploratory in nature and was designed to understand cryptographic misuses in the wild and not evaluate the tools themselves, we choose \ccsast{} due to its allowlisting approach and the inclusion of rules for JCA and BC. 
For each successfully analyzed application in our data set, \ccsast{} created a report including details of the crypto objects analyzed and the identified misuses.

\subsection{Crypto (Mis)uses - an Overview}
\label{sec:eval-cc}

\textbf{Prevalence of JCA and BC.}
We consider an application to use a crypto API if we found at least one usage of either JCA or BC, without judging the security of the respective usage yet. 
In total, we found \pgfkeysvalueof{projects_all} applications that use a crypto API. 
All of these projects use the standard Java crypto library (JCA), while \pgfkeysvalueof{projects_bc} of these projects also use the BC API. 
Within these applications, we analyzed \pgfkeysvalueof{objects_all} different crypto objects (hereafter object)\footnote{\ccsast{} uses the term \textsc{Object} to refer to any initialized Java object which interacts with a class covered by a CrySL rule~\cite{kruger2019crysl}} for JCA and BC.

\textbf{Misuses of JCA and BC.}
In the \pgfkeysvalueof{projects_all} projects, we spotted crypto misuses in \pgfkeysvalueof{misuses_all_project} and out of these \pgfkeysvalueof {misuses_bc_project} projects have misuses of both the BC and JCA library.
The remaining 180 projects only use the JCA library for crypto.
While previous studies reported 95~\% \cite{kruger2019crysl} and 96~\% \cite{gao2019negative} of applications with misuses for the JCA, 
we observe  
\pgfkeysvalueof{misuses_jca_project_perc}~\% projects with at least one misuse.
Thus, the selection criteria of so-called enterprise-driven applications~\cite{spinellis_dataset_2020} may have a slight positive impact on the number of misuses and thus security of the applications.

\textbf{Leading causes of misuses (error types):}
We identified \pgfkeysvalueof{misuses_all} misuses and those are mainly distributed over \errtype{Required Predicate}, \errtype{Incomplete Operation}, and \errtype{Constraint} errors. 
Most of the misuse reports, in total 960, are due to a \errtype{Required Predicate}, which means that composing 
multiple crypto objects seems to be challenging to get right. 
An example of this misuse is in line 4 of Listing~\ref{lst:background} where the passed key is generated insecurely.
Thus, the required predicate of the function call \code{initSign}, namely a secure key, is not fulfilled and causes a misuse. 
\errtype{Incomplete operation} errors arising out of a 
missing method call contributed to 
\pgfkeysvalueof{incomplete_total} misuses. 
For example, crypto primitives may require multiple method calls for completion, e.g., initializing, updating, and finally retrieving a hash code. 
While this is the second most frequent misuse, 
only 37.80~\% of the projects are impacted by such misuses. 
Most common in applications are misuses due to insecure parameters to functions, e.g., \textit{SHA-1} as a parameter to an initialization of a \code{MessageDigest} object.  
Overall, 80~\% of the applications contain at least one such \errtype{Constraint} error, causing in total 566 misuses.
This prevalence can be explained by the fact that the security of algorithms evolves over time, algorithms may have not been updated, and may be used for tasks beside the crypto domain. 
We observed both cases during the disclosure of our manual analysis (Sec~\ref{sec:responsible}).

\begin{table}[]
    \footnotesize
    \centering
    \begin{tabular}{lrr}
    \toprule
    JCA Class & Misuses observed & Projects affected \\
    \midrule
    \code{Cipher} & 563 & 36 \\
    \code{MessageDigest} & 472 & 113 \\
    \code{SSLContext} & 414 & 68 \\
    \code{SecretKeySpec} & 218 & 53 \\
    \code{Mac} & 161 & 34 \\
    \code{Signature} & 143 & 18 \\
    \code{KeyStore} & 120 & 44 \\ 
    \bottomrule
    \end{tabular}
    \caption{An overview of the observed misuses and affected projects for all JCA classes with more than 100 misuses.} %
    \label{tab:statsclasses}
\end{table}

\textbf{Most frequently misused crypto classes:} 
The error types discussed above describe misuses from the API perspective. 
Another interesting perspective is investigating which crypto primitives were often misused based on the classes that pertain to their implementation, as certain classes may be more prone to \textit{effective false positives}, e.g., \code{MessageDigest}, or to high-severity misuses, e.g., \code{SSLContext}. 
In the following, we will discuss all classes with more than 100 misuses. 
Table~\ref{tab:statsclasses} presents the total number of misuses and affected projects for these classes. 
While in total most of the misuses occur due to the class \class{Cipher}, 80.54~\% of the projects, have no misuse of this class at all. %
Thus, affecting only 36 projects. 
We observed the second-most misuses due to the class, \class{MessageDigest} with 472 misuses in total. %
Thus, more projects (113) have at least one misuse of the class \class{MessageDigest} than no misuse of this class.
This indicates that misuses of the class \class{MessageDigest} are widespread among many applications. %
The class \class{SSLContext} contributes to 414 misuses in total within 68 projects.
We observe fewer misuses for the class \class{SecretKeySpec} with 218 misuses in total distributed among 53 projects. 
These results reveal that only a few JCA classes are frequently misused in many of the applications. 
For our data set, we can avoid \pgfkeysvalueof{topfourclasses_perc}~\% of the misuses by fixing misuses of these four classes.

\subsection{Manual Analysis of Reports (RQ1)}
\label{sec:eval-manual}

To gain a more in-depth understanding
of common (effective) false positives, we randomly 
sampled 157 (Confidence: 99~\%, margin of error: 10~\%) misuses. 
The first four authors of this paper independently analyzed the sampled misuses, with at least two reviews per misuse. 
In case of disagreement, the reviewers in question resolved them with further code review until they reached a conclusion.  
The focus of our analysis is the precision of the report as well as the context of the misuse to answer our first research question.

Previous studies~\cite{kruger2019crysl} concentrated on measuring precision under the assumption that the rules of the static analysis are correct and based on error types that can be easily verified manually, namely \errtype{Constraint} and \errtype{Type State} errors. 
In contrast, we inspected all error types and show 
that some misuses can occur due to an incomplete or outdated rule-set or occur in a non-security context.
Thus, resulting in \textit{effective false positives}. 
We also consider such issues as a contributing factor to the overall lower precision discussed in this paper. 

\textbf{Undecided Reports.} 
We explicitly marked misuses as \textit{undecided} if it was impossible to judge their validity due to cryptic or confusing error reports. 
A common reason was that the reported insecure usage could not be identified at the location of the report, nor in the respective callers. %
In total, we observed \pgfkeysvalueof{idk} misuses which fall into this category. 
All of these misuses occur for the more complex error types, namely \errtype{Incomplete Operation}, \errtype{Required Predicate}, and \errtype{Type State}.
We assume that these error reports will be of little help to the 
user
attempting to make decisions based on them. 
For the remainder of the discussion, we will focus on the remaining 126 misuses. 
 
\textbf{True Positives.}
Our qualitative analysis revealed \pgfkeysvalueof{tp} (roughly 74~\%) true positives distributed among all previously discussed JCA classes and error types. 
By using regex, e.g., with grep, to detect usages such as MD5, \pgfkeysvalueof{tp-grep} true positives can be detected. 
Analyst or developers may consider such simple and fast tools as a prefilter of more sophisticated analyses such as \ccsast{}~\cite{kruger2019crysl} and \textit{CryptoGuard}~\cite{rahaman_cryptoguard:_2019}.  
The remaining true positives are due to an incorrect order of method calls  (\pgfkeysvalueof{tp-order}), which requires a typestate analysis, storing secrets in Java strings (\pgfkeysvalueof{tp-string}), using a default value of the JCA (\pgfkeysvalueof{tp-default}), or hard-coded credentials (\pgfkeysvalueof{tp-hardcoded}). 
The remaining 8 instances that we classified as true positives from the static analysis depends on information that can not be dissolved statically. 
True positives due to incorrect call orders and relying on defaults are to the best of our knowledge not covered in existing denylist approaches. 
Thus, showcasing the importance of a sophisticated static analysis beyond simple pattern matching. 

We observed \pgfkeysvalueof{loop} misuses because a required call is only present in a loop body where the respective loop may not be executed (Listing~\ref{lst:loop}, Line \ref{lst:l:body}). 
Thus, a misuse may be present and causing a vulnerability. 
While these cases should be analyzed to identify if a vulnerability can be triggered causing a true positive, most often these cases results into effective false positives indicating a coding smell. 
Note, that this is a known limitation of analyses such as \ccsast{}~\cite{kruger2019crysl}.

  \begin{listing}
\begin{minted}[breaklines,numbers=left,firstnumber=last,fontsize=\small,escapeinside=!!]{java}
final MessageDigest md = getMessageDigest(algo);
final byte[] buffer = new byte[4096];
int count = 0;
while ((count = inputStream.read(buffer)) > 0) { !\label{lst:l:guard}!
    md.update(buffer, 0, count); !\label{lst:l:body}!
}
return md.digest();
\end{minted}
  \caption{A \code{MessageDigest} object which only calls \code{update} when it has something to read from an \code{InputStream}.}
  \label{lst:loop}
\end{listing}

\textbf{False Positives.}
Our manual analysis resulted in 35 false positives caused by several reasons. %
One reason are incorrect API specifications. \ccsast{} relies on the correctness of the API specifications, and in case of incorrect or outdated specifications, the respective report is erroneous and leads to a false positive.
In total, we observed \pgfkeysvalueof{fp-crysl} misuses due to this reason. 
For example, the \textit{Optimal Asymmetric Encryption Padding (OAEP)} was encoded with a typo as \textit{OEAP} in the CrySL rule, causing a misuse to be reported, when the correct padding scheme \textit{OAEP} is passed.
We fixed this error in a pull request which is already merged into the main branch of the CrySL rules repository. 
Previous studies have assumed the correctness of the CrySL rules~\cite{kruger2019crysl} and thus not considered these usages as false positives. 
With this assumption, our analysis reveals a true positive rate of 84~\% that is similar to previous studies~\cite{kruger2019crysl, hazhirpasand2020java}.
In another example, the call order of functions was modeled incorrectly. 
After a discussion about this issue, a pull request fixing this problem was opened. 

Besides the correctness of the specification, the underlying static analysis can cause false positives due to imprecise modeling of the program flow. 
Our reviews revealed \pgfkeysvalueof{fp-model} false positives due imprecise modeling, e.g., wrapping of crypto objects. 
The remaining false positives are due to further challenges with respect to modeling the program statically. 

\begin{figure}
\centering
\includegraphics[width=0.8\columnwidth]{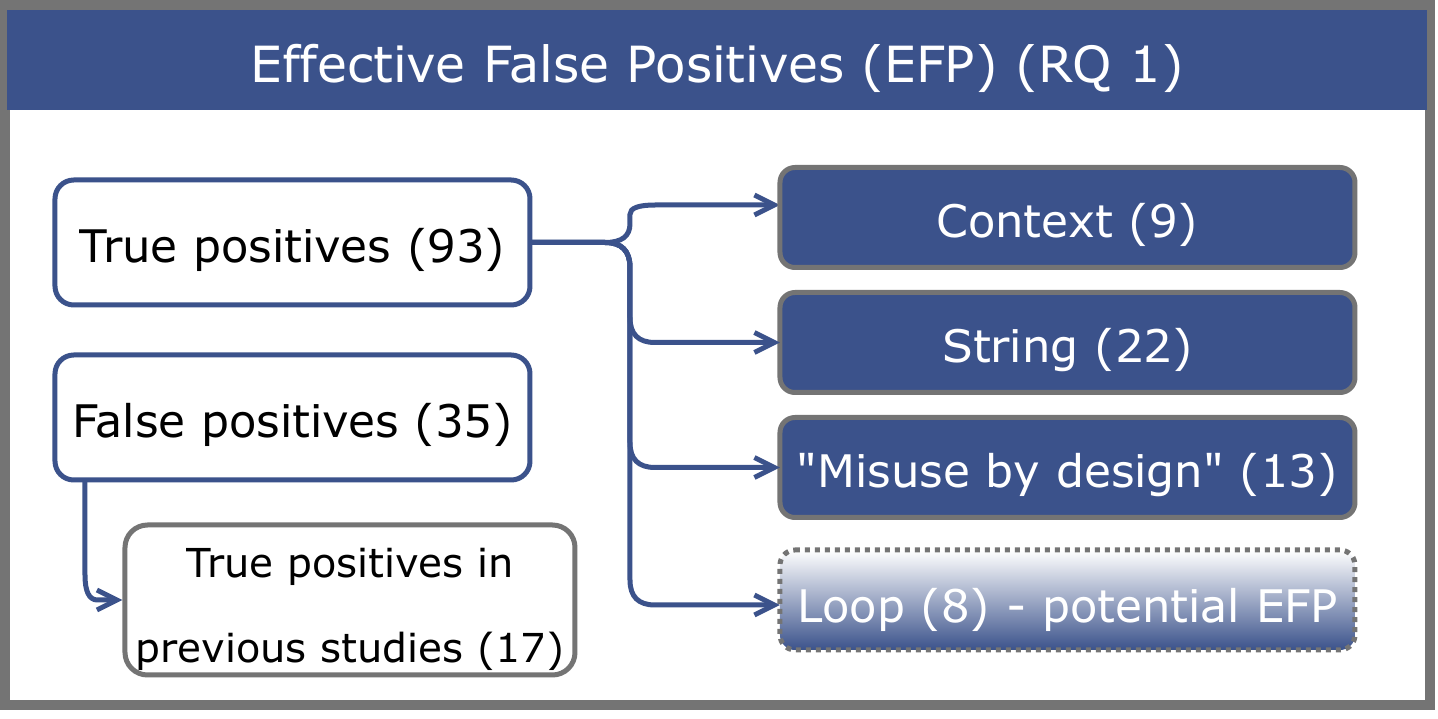}
\caption{Common \textit{effective false positives} patterns that we identified during our manual analysis.}
\label{fig:efp}
\end{figure}

\textbf{Effective False Positives.}
During our analysis, we observed that some misuses discussed previously are \textit{effective false positives} (cf. Section~\ref{def:efp}). 
Thus, the user of the analysis would not take any action, as, while they may acknowledge that it is theoretically a correct finding, it is invalid or irrelevant for the specific application. 
We provide an overview of the common patterns that we identified in the wild in Figure~\ref{fig:efp}.

Concretely, we identified \pgfkeysvalueof{non-sec-misuses} out of the 126 misuses from a non-security context. 
One example 
is the usage of \textit{MD5} in deeplearning4j\footnote{An open-source deep learning library for JVM-based languages (Stars: 12.2k, Forks: 4.9k), \url{https://github.com/deeplearning4j/deeplearning4j}}. 
Here, the \textit{MD5}-digest is used to verify that the input array %
was not modified during execution. 
Another example is in the project UAVStack\footnote{Graphical tool to monitor and analyze programs running JVM (Stars: 667, Forks: 278), https://github.com/uavorg/uavstack} where \textit{MD5}-hashes of a file are used to provide content-aware detection changes. 

The number of \textit{effective false positive} misuses due to a non-security context may appear low at first sight. 
However, one has to consider this number in relation to the overall number of reports for the specific kind, e.g., class \class{MessageDigest}.
It turns out that \pgfkeysvalueof{non-sec-misuses-md_perc}~\% of the \class{MessageDigest} misuses in our randomly selected sample fall into the category of \textit{effective false positive} due to non-security context. 
If we consider (a) that roughly 25\% of the reported \class{MessageDigest} misuses are potentially \textit{effective false positives} and (b) that this class causes by far the most of the misuses reported by previous studies~\cite{gao2019negative, kruger2019crysl, rahaman_cryptoguard:_2019}, respectively the second most frequent in our study, it becomes clear that previous reports may contain a significant number of \textit{effective false positives}. 
Further, this may significantly decrease the acceptance of these tools by developers~\cite{johnson_dev, sadowski2015Tricorder}.

Besides the context, we observed that the \pgfkeysvalueof{tp-string} true positives, which use a string instead of a byte-array, may result in \textit{effective false positives}. 
While it is possible to read user input as a byte-array, some credentials are passed by API design as a string instead of the byte-array.
Thus, the developer has no effective and easy way to fix the misuse by hand.

Our analysis revealed another potential source for \textit{effective false positives} not considered by previous studies. 
Some misuses may be intentionally introduced in the projects contained in our data set. 
For example, the static analyzer SonarSource\footnote{https://github.com/SonarSource/sonar-java (Stars: 755, Forks: 501)} includes tests, not following the typical naming scheme for test, for correct API usages and insecure crypto misuses. 
These misuses, in total 13 in our sample, within the tests are intended. 

\vspace{-0.2mm}
\begin{obs}{}{}
A non-security context, the usage of string, and "intentional misuses" are common \textit{effective false positives}.
\end{obs}

\subsection{Vulnerabilities (RQ2)}
\label{sec:eval:vuln}

\begin{figure}
    \centering
    \includegraphics[width=0.8\columnwidth]{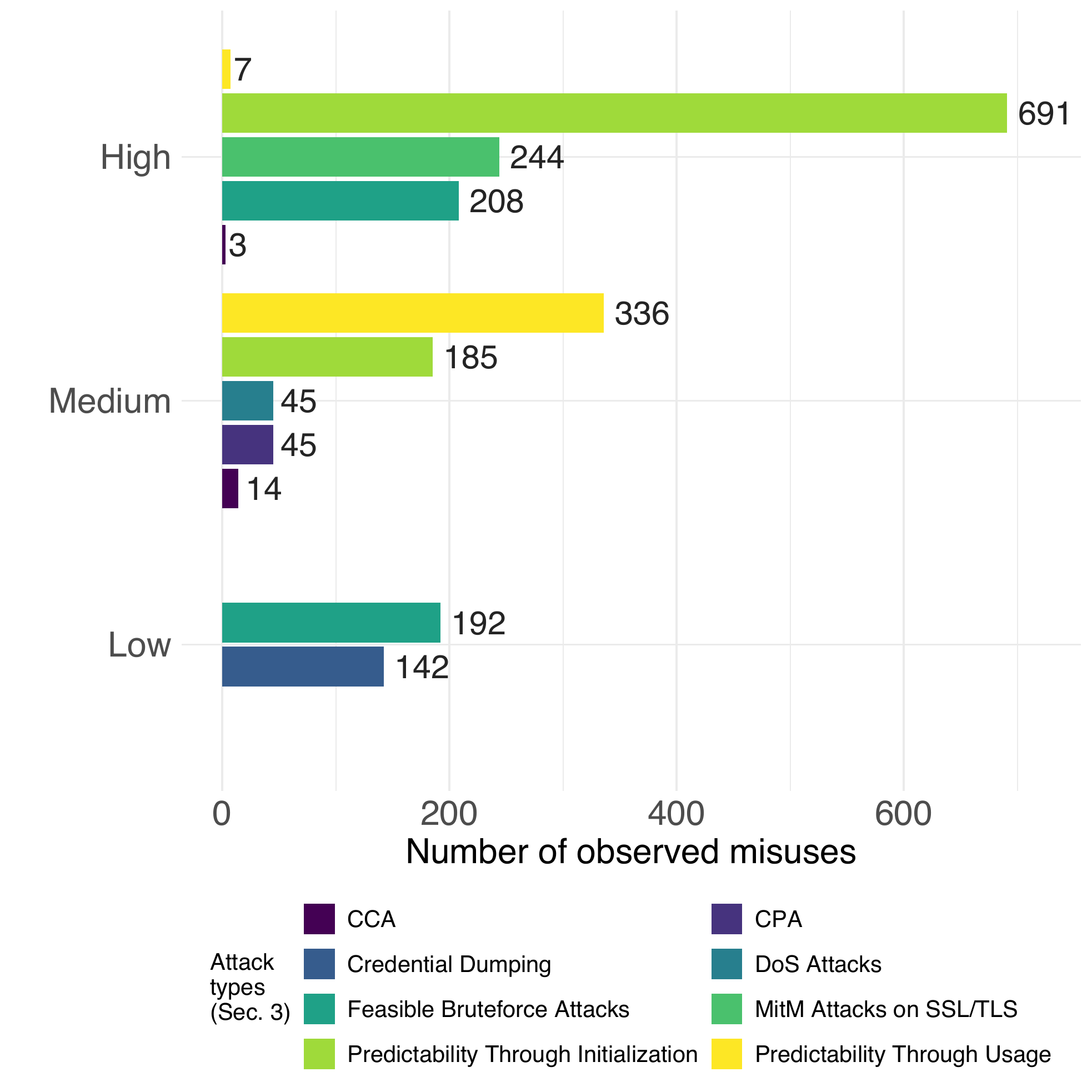}
    \caption{Number of misuses by severity for the attack types.}
    \label{fig:severity}
\end{figure}

Based upon our vulnerability model introduced in Section~\ref{sec:vulnerabilititesmodel}, we will discuss the most common vulnerabilities and their severity in this section.
Most of the misuses are due to the attack types \textit{Predictability Through Initialization} (876), \textit{Bruteforce Attacks} (400), and \textit{Predictability Through Usage} (343). 
In total, we identified 1,153 high-severity, 643 medium-severity, and 334 low-severity misuses. 
We present the number of misuses, their severity, and their relationship to the respective attack types in Figure~\ref{fig:severity}.

\textbf{High-severity vulnerabilities.}
Vulnerabilities belonging to the attack type \textit{Predictability Through Initialization} are caused by 691 misuses spread over 108 projects. 
They are due to an insecurely generated key caused by a \errtype{Required Predicate} error. 
Another source of insecurely generated keys are insecure key generation parameters, e.g., \textit{HmacSHA1} or \textit{DES}. 
These were reported for 84 misuses within 29 projects.

The usage of SSL, TLS 1.0, and TLS 1.1 is insecure and should be avoided. 
We found 141 such misuses in 61 individual projects. 
For example, 29 misuses within 20 projects are caused by using SSL. %

\textbf{Medium-severity vulnerabilities.}
Our analysis detected 336 misuses which can result in \textit{Predictability Through Usage} by observing the crypto function calls, their call order, and inputs.
For example, the analysis identifies at least one path which consumes the crypto object without adding any data, e.g., instantiating a message digest without providing an input.  
Beside the predictable computations, the most relevant medium-severity vulnerabilities are due to insecure PRNGs (total: 185) and DoS attacks caused by exceptions (total: 45). 

\textbf{Low-severity vulnerabilities.}
Most (192) of the low-severity vulnerabilities are of attack type \textit{brute-force} for ciphers, e.g., using an insecure signature algorithm or a 64-bit block cipher. 
For the attack type \textit{credential dumping}, we identified 130 low-severity misuses within 47 projects due to the use of a string instead of a byte-array for passing on secrets. 
If an attacker has control over the system, they might retrieve the secret handled as string from memory as it is immutable and cannot be cleared from memory. 
Furthermore, we found 12 misuses which use a byte array for their secrets and misses to clear the memory explicitly. 
Thus, the secret may stay in memory as for the previously discussed secrets handled as strings.  

\vspace{-0.2mm}
\begin{obs}{}{}
Nearly half the misuses (42.78~\%) are of high-severity and should be prioritized while fixing misuses. 
\end{obs}

\subsection{Responsible Disclosure.} 
\label{sec:responsible}
We informed - when possible - all projects for which we could confirm a true positive from the analysis perspective.
To avoid influencing the maintainers, we did not provide information about our judgment w.r.t \textit{effective false positives}. 
So far, we received feedback for 22 misuses from 55 reported misuses.
One project fixed the misuse that we classified as high severity and 
15 projects considered the misuse that we reported as an \textit{effective false positive}, with responses ranging over "\texttt{MD5} is used to generate test data", "the affected program component is not shipped to the customers", "the usage is secure in our setting", or "misuses in dependencies are ignored".  
One of these projects added documentation to the identified insecure usage, and initiated an internal analysis of their code for crypto misuses.
Overall, the disclosure confirms our observation that a more fine-grained assessment of misuses is important, e.g., considering a context. 
For five misuses, we received no final decision from the maintainers, e.g., they only thanked us for the report, and for one misuse we were asked to provide a fix that requires to change their API. %
Further, some discussions raised the questions which entity, e.g., the maintainers or the user of an application, is responsible for fixing a misuse.

\section{Related Work}
\label{sec:relatedWork}

Previous studies on crypto misuses have primarily focused on Android applications and identified that 88~\% to 99~\% of the applications using crypto have at least one misuse~\cite{egele_empirical_2013, muslukhov_source_2018, kruger2019crysl, rahaman_cryptoguard:_2019, piccolboni_crylogger_2021}.
Beside Android applications, Krüger et al.~\cite{kruger2019crysl} inspected maven artifacts, mostly Java libraries. 
To demonstrate the scalability of their tool \textit{CryptoGuard}, Rahaman et al.~\cite{rahaman_cryptoguard:_2019} analyzed 46 Apache projects in addition to their set of Android applications. 
All studies focused on precision from the tool's perspective, rather than understanding false positives from the developer's or security analyzer's perspective. 
An exception is an exploratory study~\cite{hazhirpasand2020java} that opened issues for some identified misuses. 
The results obtained mostly agree with the feedback we received from the disclosure as well as with our manual analysis. 
In contrast, we not only rely on the judgement of developers by manually analyzing the misuses.

Beside Java, IoT firmware, iOS applications and Python projects have been analyzed, revealing that nearly one fourth to 82~\% of the software is vulnerable~\cite{zhang_cryptorex_2019, li2015icryptotracer, feichtner_automated_2018, wicker2021Python}.
Previous research about misuses in code available from Q\&A platforms such as Stack Overflow
reveals that 71~\% to 90~\% of the posts have at least one misuse~\cite{braga_mining_2016}. %
Thus, 98~\% of the Android Apps with code copied from Stack Overflow snippets have at least one misuse~\cite{acar_comparing_2017}.

\section{Conclusion}
\label{conclusion}

In this paper, we presented a study of Java crypto misuses that focuses on understanding (effective) false positives and the severity of misuses. 
To understand common \textit{effective false positives} that arise from misuses, we manually reviewed a random sample of misuses %
and identified several \textit{effective false positives} such as the ones happening in a non-security context. 
We provide an extensive threat model covering previously undiscussed vulnerabilities such as DoS attacks in the context of crypto API misuses. %
Our threat model marks nearly half of the misuses as high-severity. 
For example, 29.05~\% of the applications still use SSL, TLS 1.0, or TLS 1.1. and may be exploitable remotely.

Overall, our study reveals several directions for future work.
While we got first insights into the differences of misuses between the JCA and BC, future research can answer the question if the BC API supports developers to write more secure code. 
Further, our manual analysis reveals several sources of \textit{effective false positives} that can be addressed in existing static analysis tools.

\section*{Acknowledgments}
We want to thank the anonymous reviewers and everyone who supported us on our journey to publish this work. 

This research work has been founded by the Deutsche Forschungsgemeinschaft (DFG, German Research Foundation) – SFB 1119 – 236615297 and – SFB 1053 – 210487104 – and by the German Federal Ministry of Education and Research and the Hessen State Ministry for Higher Education, Research and the Arts within their joint support of the National Research Center for Applied Cybersecurity ATHENE, by the LOEWE initiative (Hesse, Germany) within the emergenCITY center.

\bibliographystyle{splncs04}
\bibliography{meta/bibliography.bib}

\end{document}